\documentclass[twocolumn,showpacs,pre,superscriptaddress]{revtex4}

\begin{document}

\title{Lipid membranes with an edge}
\author{R. Capovilla}
\email{capo@fis.cinvestav.mx}\affiliation{Departamento de F\'{\i}sica,
Centro de Investigaci\'on y de Estudios Avanzados del IPN,
 Apdo. Postal 14-740,07000 M\'exico, DF, MEXICO}
\author{J. Guven}%
\email{jemal@nuclecu.unam.mx}
\affiliation{%
Instituto de Ciencias Nucleares,
Universidad Nacional Aut\'onoma de M\'exico,
Apdo. Postal 70-543, 04510 M\'exico, DF, MEXICO}
\author{J.A. Santiago}
\email{santiago@nuclecu.unam.mx}
\affiliation{%
Instituto de Ciencias Nucleares,
Universidad Nacional Aut\'onoma de M\'exico,
Apdo. Postal 70-543, 04510 M\'exico, DF, MEXICO}
\date{\today}

\begin{abstract}
Consider a lipid membrane with a free exposed edge. The energy describing 
this membrane is quadratic in the extrinsic curvature of its geometry; that
describing the edge is proportional to its length. In this note we determine 
the boundary conditions satisfied by the equilibria of the membrane on 
this edge, exploiting variational principles. 
The derivation is free 
of any assumptions on the symmetry of the membrane geometry. 
With respect to earlier 
work for axially symmetric configurations,
we discover the existence of an additional 
boundary condition which is identically satisfied in that limit.
By considering the balance of the forces operating at the edge,
we provide a physical interpretation for the boundary conditions. 
We end with a discussion of the effect of the  addition of a 
Gaussian rigidity term for the membrane. 
\end{abstract}

\pacs{87.16.Dg, 46.70.Hg}

\maketitle
  
\section{Introduction}

A closed lipid membrane is described remarkably well 
by a geometrical hamiltonian. 
This hamiltonian is constructed as a sum of the scalars, truncated at an
appropriate order, 
which characterize those features of the membrane geometry which are 
relevant. A term quadratic in the extrinsic curvature provides a measure of the 
energy penalty associated with
bending \cite{Nel.Pir.Wei:89,Pel:94,Saf:94,Lip.Sac:95,Sei:97}; 
any intrinsic tendency to bend one way and not the other
is captured by a term linear in the extrinsic curvature \cite{Hel:73}.

The shape equation determining the equilibria of this
membrane is a fourth-order non-linear elliptic PDE of the form
$\nabla^2 K+$  `$K^3$' $=0$, where $\nabla^2$ is the Laplacian on the membrane,
$K$ is the sum of the principal curvatures, and by $` K^3$' we 
mean a cubic polynomial in these curvatures \cite{Hel.OuY:87}.
Here, we would like to examine the boundary conditions which must 
supplement this equation when the membrane possesses a  free edge.
The energy cost associated with this edge is, to a good 
approximation, proportional to the exposed length. 
During the formation process, material will either be added to the edge or the
edge will heal itself so as to form closed structures.
However, unlike a simple soap film with a free edge, 
a lipid membrane does possess stable (cup shaped)
equilibria \cite{Boa.Rao:92}. 
An  
examination of such structures is a small but important step 
towards a fuller understanding 
of self assembly.   

Our primary focus will be on the boundary geometry.
We have a surface with a
boundary and a certain energy penalty associated with it, a well defined  
problem in classical field theory.
The boundary conditions are identified by demanding that the energy 
be stationary for arbitrary deformations of the edge geometry.
In distinction to earlier analytic treatments of this problem, 
we will relax the assumption that the membrane geometry be axially symmetric
\cite{Boa.Rao:92}. This is important not only for conceptual reasons.
Generally, there will be no priveleged parametrization such as that tailored to axial symmetry;
in an axially symmetric geometry the edge itself is simply a circle. 
We find that there are three boundary conditions. As we will demonstrate 
one of these conditions, involving three derivatives of the embedding function,
is satisfied identically in the axially symmetric limit.
Therefore this limit cannot be considered as a
reliable guide to the general case.

While the variational approach does capture the geometrical nature of the boundary 
conditions, the physical interpretation of these conditions  
still needs to be clarified. 
Ideally, one would like to interpret them in terms of 
the balance of forces operating at the edge. 
To do this in a way which does justice to the geometry, we identify the 
conserved Noether currents associated with the 
intrinsic translational invariance of the configuration 
 \cite{Cap.Guv:02a}.  
The three, apparently unrelated, boundary conditions are now cast in terms
of the three 
components of a single vector identity on the edge.

We finish with a discussion of the effect of a Gaussian rigidity term on 
a lipid membrane with edges. 
Whereas such a topological term does not alter the bulk shape equation, 
we show that it does modify the boundary conditions which apply to it
in a way which will have consequences in the bulk.   
This extension is relevant in topology changing processes.

The outline of the paper is as follows.
In Sect. II we consider the simple example of a surface 
tension dominated membrane. This allows us to establish our
notation and to derive the boundary conditions in a 
simple context. In Sect. III, we derive the boundary conditions
at the edge for a lipid membrane. We then specialize to 
axially symmetric configurations to compare our results in this limit 
with previous work on the subject. In Sect. IV, we consider the 
balance of the forces operating at the edge and we show
how they are related to the boundary conditions. The effect
of a Gaussian rigidity term to the membrane energy is
the subject of Sect. V, where we obtain the 
appropriate modifications in the boundary conditions.
We end with some remarks in Sect. VI.

\section{Surface tension pitted against edge tension}

It is useful to  examine first the simpler 
situation in which the membrane physics is dominated by surface tension, such as 
a soap film with a free edge. 
Let the membrane surface $\Sigma$ have an 
area $A$, with boundary ${\cal C}$ of length $L$, and 
the tension in the membrane 
bulk be a constant $\mu$,
and that on the edge $\sigma$.
The energy is then  given as a sum of two terms
\begin{equation}
F = \mu A  + \sigma L \,.
\label{eq:F}
\end{equation}
Surface tension tends to 
decrease the membrane area; line 
tension to decrease the length of the free boundary.
Without some further refinement this model does not admit stable equilibria. 
Suppose a hole is punctured in the film, then depending on its radius,
either the hole will close healing the film, or grow and 
destroy it. An unstable equilibrium clearly 
exist when the radius is tuned to coincide with a critical value
$r_c$. On dimensional ground, one world expect
$r_c \approx \mu/\sigma$. (We will ignore stability here as our interest in this model
is only as a point of reference  for a lipid membrane.)

The membrane surface $\Sigma$ is described by the  embedding ${\bf X}$ 
in three dimensional space 
$R^3$ as  
${\bf x} = {\bf X}(\xi^a)$, where ${\bf x}$ are coordinates for 
$R^3$, and $\xi^a$ coordinates for the surface ($a,b,\cdots=1,2$).
Its edge ${\cal C}$ is embedded in turn as a curve on 
$\Sigma$ as $\xi^a= Y^a(s)$ which we parametrize by its arclength
$s$.
We can now cast $F$ as
\begin{equation}
F = \mu \int_{\Sigma} d^2 \xi\; 
\sqrt{\gamma } + \sigma \oint_{{\cal C}} ds
\,.
\label{eq:FF}
\end{equation}
Here, the metric induced on $\Sigma$ is given by
$ \gamma_{ab} = {\bf e}_a\cdot {\bf e}_b $,
where ${\bf e}_a := \partial_a {\bf X}$ are tangent to the surface, $\gamma = {\rm det}\, \gamma_{ab}$, and $dA = \sqrt{\gamma} \; d^2 \xi$.
Note that we can also consider the direct
embedding of the edge ${\cal C}$ in $R^3$, via
${\bf x} = {\bf Y}(s)$, 
where ${\bf Y}  ={\bf  X}(Y^a(s))$. 
The tangent to ${\cal C}$ in $R^3$ is
equivalently expressed in either of two ways:
${\bf t}  = {\bf e}_a t^a $, 
where $t^a = \dot Y^a$; or 
${\bf t} = \dot {\bf Y}$, where a dot denotes a derivative with respect to arclength $s$.

The energy is a functional of the embedding ${\bf X}$ of 
$\Sigma$ in $R^3$. There is no need to vary the edge embedding 
$Y^a$ independently: the
$Y^a$ are fixed by the constraint that the two embeddings agree, ${\bf Y}= {\bf X} $, on ${\cal C}$. 

Equilibrium configurations are those at which
the energy Eq.(\ref{eq:F}) is stationary. 
To derive the equations describing the equilibrium configurations of a 
membrane described by this model, we first consider a 
variation of the embedding ${\bf X}$ of the membrane
${\bf X}\to {\bf X} + \delta {\bf X}$.
We let ${\bf n}$ denote the 
unit normal to the surface $\Sigma$. We decompose the displacement with
respect to the
spatial basis adapted to $\Sigma$, $\{{\bf e}_a , {\bf n}\}$,  as,
\begin{equation}
\delta {\bf X} = \Phi^a \, {\bf e}_a + \Phi\, {\bf n}\,. \label{eq:delta}
\end{equation}
We now have that the induced metric varies according to (see {\it e.g.} 
 \cite{Cap.Guv:02a})
\begin{equation}
\delta_X \gamma_{ab} = 2 K_{ab}\Phi 
+ \nabla_{a}\Phi_{b} +\nabla_b \Phi_a \,,\label{eq:delgam}
\end{equation}
where $K_{ab}$ denotes the extrinsic curvature tensor,
\begin{equation}
K_{ab}=  {\bf e}_b \cdot  \partial_a\, {\bf n}\,,
\end{equation}
and $\nabla_a$ is the covariant derivative on $\Sigma$ compatible with $\gamma_{ab}$.
The derivative terms in the variation of $\gamma_{ab}$ are its Lie derivative along the tangential 
vector field, $\Phi^a$. The variation of $A$ is
\begin{eqnarray}
 \delta_X A
%&=&   
%\int_{\Sigma} dA \,\gamma^{ab} \delta_X \gamma_{ab}
%\nonumber\\
&=& \int_{\Sigma} dA\; 
\left[  K\; \Phi
+ \nabla_a\; \Phi^a \right] \nonumber \\
&=& \int_{\Sigma} dA\; 
K \; \Phi + 
\oint_{{\cal C}} ds\;  l^a \; \Phi_a\,. 
\label{eq:16}
\end{eqnarray}
The mean extrinsic curvature, 
$K = K_{ab} \gamma^{ab}$. The second line follows from the preceding one 
using Stoke's theorem.
Here $ l^a$ is the outward pointing normal to ${\cal C}$ 
on $\Sigma$.
Only the normal projection $\Phi$ of the variation
plays a role in determining the bulk equilibrium of the membrane.
This is true generally, regardless of the model. 
In this particular model,
however, there is no boundary 
term associated with the bulk normal displacement $\Phi$. As we will see, this is not generally 
true --- a happy accident when the energy is truncated at the
area term. On the other hand, the tangential bulk variation $\Phi^a$ always  gives only a boundary 
term. This is a consequence of the fact that a tangential deformation 
corresponds in the bulk to an infinitesimal reparametrization of the surface. 
There is however a physical displacement of the boundary. 
In fact, the boundary contribution to Eq.(\ref{eq:16}) is 
easily identified as the
change in the surface area of $\Sigma$ under a normal deformation of 
its boundary, $\delta Y^a =  (l^b \, \Phi_b )\, l^a$, which at each point 
is directed along the tangent plane of $\Sigma$ at that point.
The projection of $\Phi^b$ onto the edge ${\cal C}$ 
itself, $ t^a \Phi_a$, does not contribute.

Let us turn now to the  variation of the edge embedding ${\bf Y}$ induced
by the bulk variation $\delta {\bf X}$. It can 
be decomposed with respect to a basis adapted to both embeddings,
${\bf X} $ and ${\bf Y}$, given by $\{{\bf t},{\bf l},{\bf n}\}$, 
where ${\bf l}= {\bf e}_a  l^a$. Thus at the edge we set 
\begin{equation}
\delta {\bf Y} = \phi \;{\bf t} + \psi \;{\bf l}  + \Phi \;{\bf n}
\,,
\end{equation}
where the edge and bulk components are identified by continuity, $\psi= l^a \; \Phi_a$ and $\phi = t^a \; \Phi_a$. 
Modulo a divergence associated with a reparametrization of the 
boundary which involves the tangential component $\phi$ that we can
safely discard, we have for the variation of the infinitesimal 
arclength
\begin{equation}
\delta_Y \, ds  = ds \; \left( 
\kappa \, \psi + K_\parallel \, \Phi \right)
\,,
\label{eq:delZ}
\end{equation}
where we have used the fact that
\begin{equation}
\dot{\bf t} = - \kappa {\bf l} - K_\parallel {\bf n}\,.
\label{eq:dt}
\end{equation}
Here
$\kappa$ is the geodesic curvature of ${\cal C}$ associated with its 
embedding in $\Sigma$, and we have defined 
 $K_\parallel = K_{ab} t^a t^b$. The unconventional minus sign in the first
term of Eq. (\ref{eq:dt}) comes about because ${\bf l}$ is the {\it outward}
normal to ${\cal C}$ on $\Sigma$, {\it i.e.} $\dot{t}^a = - \kappa \; 
l^a$.

The corresponding deformation in $L$ is then given by 
\begin{equation}
 \delta_Y L
= \oint_{{\cal C}} ds  \, \left( 
 \kappa \, \psi + K_{\parallel}\, \Phi
\right)\,.
\label{eq:bc}
\end{equation}
Summing the two contributions (\ref{eq:16}) and (\ref{eq:bc}) 
to the variation of the 
energy $F$, 
as given by (\ref{eq:F}), we find
\begin{equation}
\delta_X F =
\mu \int_{\Sigma} dA\, 
K \; \Phi + \oint_{{\cal C}} ds\, 
\left[(\mu + \sigma\, \kappa ) \psi 
+ \sigma \,K_{\parallel} \, \Phi\right]\,.
\end{equation}
The bulk equilibrium is 
a minimal surface unaffected by the boundary, satisfying 
$K=0$. On the boundary, the projections along the normals to the edge,
$\psi$ and $\Phi$, represent 
independent deformations, so that stationarity of $F$ requires 
the vanishing of the 
corresponding coefficents. We thus read off the two boundary conditions:
\begin{eqnarray}
\sigma \,\kappa + \mu &=&0\,,\label{eq:s1}\\
\sigma \, K_{\parallel} &=&0\,.
\label{eq:s2}
\end{eqnarray}
The first tells us that the geodesic curvature 
of the edge as embedded in the membrane is 
constant. The second simply enforces the vanishing of
$K_{\parallel}$ at the edge. Note that the
completeness of the 
basis $\{{\bf t},{\bf l}\}$ of tangent vectors on $\Sigma$ at ${\cal C}$,
$\gamma^{ab} =  t^a t^b + l^a l^b $, permits us 
to express the mean curvature at the edge as 
$K = K_\parallel+ K_\perp$, where $K_\perp = K_{ab} l^a l^b$.
Thus modulo the bulk equilibrium $K=0$, 
the boundary condition (\ref{eq:s2}) can be alternatively expressed as $K_{\perp} =0$.
The only potentially non vanishing component of $K_{ab}$ on the edge is the off-diagonal component, $K_{\parallel\perp} = t^a l^b K_{ab}$.

For this particular model our approach has been  heavyhanded;
the boundary conditions we have written down
are an elaborate way to express the simple vector identity
\begin{equation}
\sigma \dot {\bf t} = \mu {\bf l}\,,
\label{eq:B'}
\end{equation}
which equates the change in the tension over the interval
$\Delta s$ along the edge, $\sigma \Delta {\bf t}$,
to the force due to surface tension acting on the edge, $\mu {\bf l} \Delta s$. The apparent mismatch in counting (three versus two) is 
accounted for by noting that the projection of Eq.(\ref{eq:B'}) along 
${\bf t}$ is an identity. For higher order models, as we will see, this projection will not be vacuous.

Note that had we $N$ sheets conjoined on a single edge, Eq.(\ref{eq:B'}) gets 
modified in an obvious way:
\begin{equation}
\sigma \dot {\bf t} = \mu \sum_{i=1}^N {\bf l}_i \,,
\label{eq:B''}
\end{equation}
where ${\bf l}_i$ is the vector normal to the edge which 
is tangent to the $i^{\rm th}$ sheet. Eq.(\ref{eq:B''}) provides a 
generalization of the Neumann rule for soap bubble clusters at a Plateau
border \cite{Wea.Hut:99}
to accomodate line tension on the edge.
A simple application is considered in \cite{Cap.Guv:02c}.

\section{Lipid membrane with an edge}

A lipid membrane is modeled by a phenomenological energy 
quadratic in the extrinsic curvature of the surface.
Let us write this as 
\begin{equation}
F_b = \int_\Sigma dA \; {\cal F} (\gamma^{ab}, K_{ab})\,,
\end{equation}
{\it i.e.} ${\cal F}$ depends at most on the extrinsic curvature, and
not, for example, on its derivative $\nabla_a K_{bc}$.
In particular, we will focus  on the model 
described by the Helfrich energy density
\begin{equation}
{\cal F} = \alpha (K-K_0)^2 + \mu \,.
\label{eq:rigid}
\end{equation}
The spontaneous curvature $K_0$ is a constant, as is the bending rigidity
$\alpha$. The  constant $\mu$ is 
interpreted here as the Lagrange multiplier implementing the 
constraint on the membrane area. 
We will discuss the addition of a Gaussian rigidity term in the next section.

The energy of the bulk and the edge is
\begin{equation}
F = F_b + \sigma L\,.
\end{equation}
 
The shape equation describing the equilibrium in the bulk, which is
derived from the extremization of the energy (\ref{eq:rigid})
\begin{equation}
\alpha [ -2 \nabla^2 K +  2 (K- K_0) {\cal R}  + (K_0{}^2 - K^2 )  K
]  +\mu \, K = 0\,,
\label{shape}
\end{equation}
is well known \cite{Hel.OuY:87}. The structure of this equation has been 
discussed in detail elsewhere \cite{Cap.Guv:02a}, where a novel derivation  
is also provided. 
The scalar curvature ${\cal R}$ appearing in Eq.(\ref{shape}) 
is related to the extrinsic
curvature through the Gauss-Codazzi equation,
\begin{equation}
{\cal R} = 
K^2 - K_{ab} K^{ab} 
\,.\label{eq:GC}
\end{equation}

Under a tangential deformation of the surface,
$\delta_{\parallel}{\bf X}  = \Phi^a {\bf e}_a$, 
the energy density transforms as 
a divergence which is transferred to the boundary,
\begin{equation}
\delta_{\parallel} F_b = \oint_{\cal C} ds\, {\cal F} \, l_a \, \Phi^a \,.  
\label{eq:bb}
\end{equation}
This is because the local scalar energy density ${\cal F}$ transforms as
\begin{equation}
\delta_{\parallel} {\cal F} = \Phi^a \, \partial_a {\cal F}\,.
\end{equation}
The details of ${\cal F}$ do not enter.
Note that Eq.(\ref{eq:bb})  agrees with the corresponding expression for the area 
with ${\cal F}=1$.
As before, this boundary term induces a source
into the boundary Euler-Lagrange equation.
For an edge with a line tension $\sigma$, we get the first boundary 
condition, due to a deformation along the normal ${\bf l}$, $\psi$,
\begin{equation}
\sigma \kappa + {\cal F} =0\,,
\label{eq:b1}
\end{equation}
where we have used Eqs. (\ref{eq:bc}) and (\ref{eq:bb}).
This should be compared with Eq.(\ref{eq:s1}) to which it reduces if ${\cal F}=\mu$, a 
constant. This boundary condition relates the
geometry of the edge to the extrinsic curvature of
the membrane evaluated at the boundary.

We now examine a normal deformation of the surface 
$\Sigma$, $\delta_\perp
{\bf X} = \Phi \; {\bf n}$.
The shape equation (\ref{shape}) determining the 
local membrane equilibrium is obtained 
by demanding that the  energy be stationary 
with respect to normal deformations of $\Sigma$ which may or may not vanish on the boundary. 
As such this equation cannot be affected by
the addition of a boundary. To determine the 
boundary conditions we need to extend the support of the 
variation to include the boundary. We have
that the normal variation of the bulk energy can be written as
\begin{equation}
\delta_{\perp} F_b =
\int dA \left[ {\cal F} K\Phi + 
2\, \Gamma_{ab}  K^{ab} \Phi
+ {\cal F}^{ab}
\delta_{\perp} K_{ab} \right]\,,
\label{dFp}
\end{equation}
where $\Gamma_{ab} =
\partial {\cal F}/\partial \gamma^{ab}$ and ${\cal F}^{ab}=
\partial {\cal F}/\partial K_{ab}$.
The boundary term we wish to identify in 
$\delta_{\perp} F_b$ originates in the $\delta_\perp K_{ab}$ term
in this expression. We recall that the  extrinsic curvature transforms as follows
under a normal deformation of $\Sigma$ (see {\it e.g.}
\cite{Cap.Guv:02a}):
 \begin{equation}
\delta_{\perp} K_{ab} =
-\nabla_a \nabla_b \Phi +
K_{ac} K^c{}_{b} \Phi
\,.
\end{equation}
We thus have that
\begin{equation}
\delta_{\perp} F_b =
\int dA \left[ {\cal E}\Phi 
+
\nabla_a\Big(\Phi \nabla_b {\cal F}^{ab} 
- {\cal F}^{ab}
\nabla_b \Phi\Big)\right]
\,,
\label{dFp1}
\end{equation}
where we have defined the Euler-Lagrange derivative
\begin{equation}
{\cal E} = (- \nabla_a\nabla_b + K_{ac} K^c{}_{b})
{\cal F}^{ab} 
+ {\cal F} K + 
2\, \Gamma_{ab} K^{ab} \,.
 \end{equation}
Thus, modulo the bulk shape equation, ${\cal E}=0$, we get
that the boundary contribution is
\begin{equation}
\delta_{\perp} F_b =
\oint_{\cal C} ds \, l_a \left[
\Phi \nabla_b {\cal F}^{ab} - {\cal F}^{ab}\nabla_b \Phi\right]\,.\label{eq:deli}
\end{equation}
The terms proportional to 
$\nabla_a \Phi$ and $\Phi$ are not independent:
the projection of $\nabla_a\Phi$ along the edge is 
completely determined once $\Phi$ is specified on ${\cal C}$.
To decompose $\delta_\perp F$ 
into two independent parts we proceed as follows:
we first decompose $\nabla_a\Phi$ into its normal and tangential parts 
with respect to ${\cal C}$, 
\begin{equation}
\nabla_a \Phi =  l_a \,  \nabla_\perp \Phi
+t_a \dot\Phi\,,
\label{eq:ddc9}
\end{equation}
where we have defined $\nabla_\perp = l^a\nabla_a$.
We now perform an integration by parts on the $\dot \Phi$ term to obtain 
for the second term on 
the right hand side of Eq.(\ref{eq:deli}) 
\begin{eqnarray}
\oint_{\cal C} &ds& \, l_a 
{\cal F}^{ab} \nabla_b \Phi  = \nonumber \\
\oint_{\cal C} &ds&\,
\left[ l_a  l_b {\cal F}^{ab} \nabla_\perp \Phi - \Phi {d\over ds }\Big(
 l_a {\cal F}^{ab} 
t_b\Big) \right]\,,
\label{eq:delii}
\end{eqnarray}
where we have discarded a total derivative term with respect to arclength.
In this way we succeed in isolating the independent normal variations at the 
boundary, the coefficients of 
$\Phi$ and $\nabla_\perp \Phi$.

From Eqs. (\ref{eq:bc}), (\ref{eq:deli}), (\ref{eq:delii}), we
obtain that the total boundary contribution of the normal variations
is 
\begin{eqnarray}
\delta_\perp &F_b& = \oint_{{\cal C}} ds \; \Big\{
- l_a  l_b {\cal F}^{ab} 
 \nabla_\perp \Phi \nonumber \\
+ &\Big[& l_a  \nabla_b {\cal F}^{ab} +
{d\over ds }\Big(
 l_a {\cal F}^{ab}
t_b\Big) + \sigma K_\parallel \Big] \Phi
\Big\}\,,
\end{eqnarray}
so that we can immediately read off the 
two boundary conditions which supplement Eq. (\ref{eq:b1}), 
\begin{eqnarray}
 l_a\nabla_b {\cal F}^{ab}
+ {d\over ds} \Big(
 l_a {\cal F}^{ab} t_b\Big)
+ \sigma K_{\parallel} &=&0\,,\label{eq:b2}\\
 l_a l_b {\cal F}^{ab} &=&0\,.\label{eq:b3}
\end{eqnarray}
The first is of third order in derivatives of the embedding functions.
This is consistent with the fact that the shape equation (\ref{shape}) is 
of fourth order.  
Using the decomposition of the covariant derivative
(\ref{eq:ddc9}), it can be written in the alternative form
\begin{equation}
 l_a l_b \nabla_\perp {\cal F}^{ab} 
+ 2 {d\over ds} \Big(
 l_a {\cal F}^{ab} t_b\Big) 
+ \kappa (l_a l_b - t_a t_b ) {\cal F}^{ab}
+ \sigma K_{\parallel} = 0\,.
\label{eq:b2a}
\end{equation}

In the case of a membrane described by the Helfrich Hamiltonian  
(\ref{eq:rigid}) with an edge the third 
boundary condition (\ref{eq:b3}) implies
\begin{equation}
K=K_0\,, 
\label{k0}
\end{equation}
on the edge --- the rigid membrane necessarily has a constant mean curvature 
at the edge equal to its spontaneous value. 
This is entirely independent of the tensions, $\mu$ or $\sigma$, or of the 
rigidity modulus. If $K_0=0$, the membrane is minimal at its edge.

The second boundary condition (\ref{eq:b2}) is of Robin type.
For any ${\cal F}$ which is a function only of $K$, we
have that
${\cal F}^{ab}\propto \gamma^{ab}$, 
so that the middle term in (\ref{eq:b2}) vanishes,
\begin{equation}
 l_a {\cal F}^{ab}t_b
=0\,,
\end{equation}
and the boundary condition reduces to
\begin{equation}
2 \alpha \,\nabla_\perp K 
+ \sigma K_{\parallel} =0\,.
\label{nk}
\end{equation}
This equation determines the normal derivative of $K$ in terms of 
the component of the extrinsic curvature tangent to the edge. 
It does not involve the surface tension $\mu$.
We emphasize that its existence seems to have gone unnoticed.

The first boundary condition, Eq.(\ref{eq:b1}), together 
with Eq.(\ref{k0}), implies that on the edge 
\begin{equation}
\sigma  \kappa + \mu=0\,.
\label{eq:kon}
\end{equation}
The geodesic curvature of a loaded boundary 
is completely fixed by the ratio of the tensions  in exactly the same 
way as in the previous section for soap bubbles, see Eq. (\ref{eq:s1}) .
If $\mu=0$ 
the edge is necessarily a geodesic of the 
bulk geometry. 

If the line tension on the boundary vanishes, $\sigma=0$,  
the consistency of Eq.(\ref{k0}) with (\ref{eq:b1}) requires that $\mu=0$ also.
Furthermore Eq.(\ref{nk}) implies $\nabla_\perp K =0$ on the boundary. 
But the unique solution satisfying the 
two boundary conditions $K=K_0$ and $\nabla_\perp K =0$ is $K=K_0$ everywhere.
One way to see this is to construct the 
Gaussian normal coordinates adapted to the edge, $(l,s)$, where
$l$ is the length of the geodesic which intersects the edge normally.
With respect to this system of coordinates, the Laplacian
assumes the form $\nabla^2 = \partial_l^2 + \kappa \partial_l + \partial_s^2$
in the neighborhood of the edge. Thus, 
modulo the boundary conditions, $\nabla^2 K = \partial_l^2 K$ on the edge.
But Eq.(\ref{shape}) implies that $\nabla^2 K=0$ there so that $\partial_l^2 K$ and all
higher derivatives vanish.
If $K$ is analytic in
$l$, then $K=K_0$. If $\mu\ne 0$, there is no such constraint.
The geometry is very severely constrained by the boundary conditions.

Let us now examine an axially symmetric membrane with an axially symmetric 
edge. With respect to cylindrical polar cooordinates $\{\rho,z,\varphi\}$
on 
$R^3$, the membrane is described by
$\rho = R(\ell)$, $z = {\cal Z}(\ell)$
where ${\cal Z}'{}^2 + R'{}^2  = 1$.
$\ell$  is the arc length along a
a curve with fixed $\varphi$, 
and the primes denote a derivative with respect to $\ell$.
The intrinsic geometry of $\Sigma$ is described by the 
line element 
\begin{equation}
d\tau^2= d \ell^2 + R^2(\ell) d\varphi^2\,,
\label{eq:metcyl}
\end{equation}
We can write the extrinsic curvature in a form consistent with 
axial symmetry as
\begin{equation}
K_{ab}= \ell_a \ell_b K_\ell  + (\gamma_{ab}-\ell_a \ell_b)K_R\,,
\end{equation}
where $K_l$ and $K_R$ are two spatial scalars which we identify
as the principal curvatures of the embedding of $\Sigma$ in $R^3$,
 $ \ell^a$ is the outward pointing unit normal to the circle 
of fixed $\ell$, $\ell^a=(1,0)$.
The mean curvature is $K = K_\ell + K_R$.
To evaluate the principal curvatures,
it is convenient to define
$\Theta$ as the angle which the tangent to a curve of fixed $\varphi$ makes with the 
cylindrical radial direction:
\begin{equation}
{d {\cal Z}\over d R} = \tan \Theta\,.
\end{equation}
We then have ${\cal Z}'=\sin\Theta$, and
$R'=\cos\Theta$, so that the principal curvatures are 
\begin{equation}
K_\ell = \Theta'\,, \quad \quad 
K_R = {\sin \Theta\over R}\,.
\end{equation}
Axial symmetry implies that the fourth order 
shape equation can be integrated to provide a third order 
equation for $R$ as a function of $\ell$. 
It has been shown elsewhere (\cite {Zhe.Liu:93,Boz.Sve.Zek:97}, see also
\cite{Cap.Guv:02a}) that 
this equation takes the form
\begin{widetext}
\begin{equation}
-2 \alpha \cos\Theta \left(\Theta' + {\sin \Theta\over R} \right)' 
+ 
\alpha (\Theta' + {\sin \Theta\over R}) 
(\Theta' - {\sin \Theta\over R}) \sin\Theta + 2\alpha  K_0  {\sin^2\Theta\over R}  - 
(\mu + \alpha K_0^2) \sin\Theta
= 0\,.
\label{eq:axeq}
\end{equation} 
\end{widetext}
If the boundary ${\cal C}$ is also axially symmetric so that it
coincides with a fixed value 
of $\ell$ then $ l^a=\ell^a$, 
$K_{\parallel} = K_R$, $K_\perp = K_\ell$ and 
$K_{\parallel\perp}=0$. It is simple to show that $\kappa = - R'/R$.
We thus have for the boundary conditions, Eqs.(\ref{k0}) and 
(\ref{eq:kon}), 
\begin{equation}
\Theta' + \sin \Theta/R = K_0\,, \quad \quad \sigma R' = \mu R\,.
\label{eq:axbc}
\end{equation}
The remaining boundary condition, Eq.(\ref{nk}),
of third order in derivatives appears to present a problem: 
a third order ODE does not admit third order boundary conditions.
The inconsistency, however, is only apparent:
on the boundary, the shape equation
Eq.(\ref{eq:axeq}) itself reproduces,
modulo Eqs.(\ref{eq:axbc}), the troublesome boundary condition 
Eq.(\ref{nk}).
Our analysis is thus completely consistent with the axially 
symmetric analysis of \cite{Boa.Rao:92}
where the boundary conditions Eqs.(\ref{eq:b1}) and 
(\ref{eq:b3}) are derived. 
It is worth stressing, however, that potential pitfalls 
of using the axially symmetric problem as  
a guide to the more general problem.  The boundary condition (\ref{eq:b2}) 
is a non-trivial constraint on the geometry which is not 
already encoded in the 
shape equation for non-axially symmetric configurations.

\section{Balance of forces}

In this section, we consider the balance of the forces operating at the
edge. This provides the missing intuition on the physical origin of the
boundary conditions we have derived in the previous section.

Consider a point on the edge. In equilibrium,
the tension ${\bf g}$ must satisfy 
\begin{equation}
\dot{\bf g} = {\bf f}^a l_a\,.
\label{gf}
\end{equation}
Here ${\bf f}^a$ is the membrane stress tensor so that 
${\bf f}^a l_a$ is the `surface tension' acting on the edge due to 
unbalanced stresses in the bulk at its boundary.
In \cite{Cap.Guv:02a}, it was shown that the bulk stress tensor 
for the model defined by the Helfrich energy (\ref{eq:rigid}) can be expressed in the  form
 \begin{eqnarray}
{\bf f}^a &=& 
[ 2\alpha K (K^{ab} - {K\over 2} \gamma^{ab}) - 2 \alpha K_0 (K^{ab} - K
\gamma^{ab}) \nonumber \\
&-&(\mu + \alpha K_0^2)  \gamma^{ab} ]  {\bf e}_b 
- 2 \alpha \nabla^a K {\bf n}\,.
\end{eqnarray} 
Thus its projection along the normal to the edge $l^a$ is
\begin{eqnarray}
{\bf f}^a l_a  &=& 
\left\{2\alpha (K - K_0)  K_\perp - \alpha (K-K_0)^2 - \mu
\right\} {\bf l} \nonumber \\
 &+& 2\alpha (K - K_0) K_{\parallel\perp} \,
{\bf t}
- 2 \alpha \nabla_\perp K {\bf n}\,.
\end{eqnarray}
In addition, as we have seen in Sect. II, 
\begin{equation}
{\bf g} = -\sigma {\bf t}\,.
\label{eq:gt}
\end{equation}
Using Eq. (\ref{eq:dt}) for $\dot{\bf t}$, 
we read off the three components of Eq.(\ref{gf}), 
\begin{eqnarray}
\sigma \kappa &=& 2\alpha (K - K_0)  K_\perp - \alpha (K-K_0)^2 -
\mu\,, \label{n1}\\
\sigma K_\parallel &=& - 2\alpha \nabla_\perp K \,,\label{n2}\\
0 &=& 2 \alpha (K - K_0) K_{\parallel\perp} \label{n3}\,,
\end{eqnarray}
respectively along ${\bf l}$, ${\bf n}$, and ${\bf t}$.
The condition (\ref{n2}) coincides with the boundary condition (\ref{nk}). 
If $K_{\parallel\perp}\ne 0$, Eq.(\ref{n3}) implies that $K=K_0$. 
The remaining boundary condition (\ref{n1}) then coincides with a linear combination
of the  boundary conditions (\ref{k0}) and (\ref{eq:kon}). 
In the axially symmetric geometry, however, $K_{\parallel\perp}$ does vanish so that
(\ref{n3}) does not imply $K=K_0$ as it stands.  
One needs then to appeal to the integrated shape equation (\ref{eq:axeq}), which 
together with (\ref{n1}) and (\ref{n2}) reproduce $K=K_0$.

We thus have identified a very simple (if heavily disguised) 
physical interpretation of the boundary conditions.
In particular, in this approach, the boundary condition
$K = K_0$ emerges as the vanishing of the stress induced by the bulk along the edge.
Note that the variational approach did not rely on the identification of 
projections. Indeed, the boundary condition corresponding to the projection 
along ${\bf t}$ was originally identified
by demanding stationary energy for independent boundary variations of $\nabla_\perp\Phi$.

We also note that the form of Eq.(\ref{gf}) implies the integrability condition,
\begin{equation}
\oint_{\cal C} ds \, {\bf f}^a l_a =0
\end{equation}
on the edge. The existence of these three extremely non-trivial conditions is 
far from obvious in our 
previous approach.

One can say more. Take the equation, $\nabla_a {\bf f}^a =0$, describing 
the conservation of the stress tensor. dot it into ${\bf X}$ and integrate over the 
membrane surface. We get

\begin{equation}  
\int dA\, \nabla_a \, ({\bf X}\cdot {\bf f}^a) = \int dA\, {\bf e}_a\cdot {\bf f}^a\,.
\end{equation}
Working on the right, we have
\begin{eqnarray}  
\int dA\, \nabla_a \, ({\bf X}\cdot {\bf f}^a) &=& \oint ds\, {\bf X} \cdot l_a {\bf f}^a\nonumber\\
&=& \oint ds\, {\bf X} \cdot \dot {\bf g}\nonumber\\
&=&- \oint ds\, {\bf t} \cdot {\bf g}\nonumber\\
&=&\sigma L \,,
\end{eqnarray}
where we have used Eq.(\ref{gf}) on the second line,
as well as Eq.(\ref{eq:gt}) on the last line.
On the other hand, if we write ${\bf f}^a = f^{ab} {\bf e}_b + f^a {\bf n}$, then

\begin{equation}
 \int dA\, {\bf e}_a\cdot {\bf f}^a = \int dA\, f^a{}_a\,.
\end{equation}
We note that the bending energy $\int dA\, K^2$ is a conformal invariant,
and so does not contribute to the trace $f^a{}_a$.
We have $f^a{}_a =  2(\alpha K_0 (K  - K_0) - \mu)$,
so that
\begin{equation}
2\mu A  -  2\alpha K_0 \int dA\, (K-K_0) + \sigma L =0\,.
\end{equation}
This condition is useful for identifying the sign associated with the mulupliers. 
For example, if $K_0=0$, it is clear that $\mu$ is necessarily negative.

\section{Gaussian rigidity}

The geometrical scalars we can construct with dimension
$[\rm Length]^{-2}$ are ${\cal R}$, $K^2$ and $K_{ab}K^{ab}$. 
%These are related by the Gauss-Codazzi and Codazzi-Mainardi equations,
%(\ref{eq:GC}) and
%\begin{equation}
%\nabla_a K^{ab} - \nabla^b K=0\label{eq:CM}
%\end{equation}
%respectively.
The Gauss-Codazzi equation (\ref{eq:GC}) 
tells us that the three scalar invariants
${\cal R}$, $K^2$ and $K_{ab} K^{ab}$ are not independent.
In addition, the Gauss-Bonnet functional
\begin{equation}
I =\int_\Sigma dA \,{\cal R}
\end{equation}
is a topological invariant if the membrane is closed.
More generally for an open membrane,
\begin{equation}
\label{EH}
I = \int_{\Sigma}
dA\, {\cal R} 
+ 2 \oint_{{\cal C}} ds\, 
\kappa
\end{equation}
is a topological invariant. 
A consequence is that if a Gaussian rigidity term is included in the 
energy a line rigidity $\oint ds \kappa$
is necessarily induced along its boundary.

To obtain the variation of the Gaussian term,
we need to know how the scalar curvature
varies. Its tangential deformation is straightforward; to 
determine its normal deformation, we exploit Eq.(\ref{eq:GC})
and with it the technology developed Sects. II and III. 

Consider now a Gaussian rigidity addition to the  bulk energy, 
so that
\begin{equation}
F= F_b + \beta \,\int dA\,{\cal R}\,,
\end{equation}
Whereas the bulk
shape equation is unmodified, all three boundary conditions
are changed:
\begin{eqnarray}
\alpha (K-K_0)^2 + \beta {\cal R} + \mu + \sigma \kappa &=& 0\,,
\label{bb1}\\
 2\alpha \nabla_\perp K - 2\beta   \dot{K}_{\parallel\perp}
 +  \sigma K_{\parallel} &=&0 \,,\label{bb2}\\
 \alpha (K-K_0 ) + \beta K_{\parallel} 
&=&0\,.\label{bb3}
\end{eqnarray}
We note that the Gauss-Codazzi equation (\ref{eq:GC})
allows us to express ${\cal R}$ in terms of the 
projections of $K_{ab}$ with respect to the edge,
${\cal R}= 2 (K_{\parallel} K_{\perp} -
K_{\parallel\perp}{}^2 )$. Eq.(\ref{bb1}) is quadratic in the 
extrinsic curvature. Eqs.(\ref{bb2}) and (\ref{bb3}) by constrast are linear
relationship between $K_{\perp}$ and $K_{\parallel}$. 

Note that, unlike the case of the pure Helfrich model,
the central term in Eq.(\ref{bb2}) does not vanish in general.
However, it does vanish in an axially symmetric geometry (with axially
symmetric
edge), $K_{\parallel\perp}=0$. More generally, we have
the integral statement

\begin{equation}
 \oint_{\cal C} ds \left[ 2\alpha \nabla_\perp K
 + \sigma  K_{\parallel}\right]=0\,.
\end{equation}
In an axially symmetric geometry one can check 
that, modulo the lower order boundary conditions (\ref{bb1}) and
(\ref{bb3}),
Eq.(\ref{eq:axeq}) reproduces Eq.(\ref{bb2}) on the boundary. 

Let us consider now the balance of the forces in this case. 
The Gaussian term makes no contribution to ${\bf f}^a$ \cite{Cap.Guv:02a}.
Naively reinvoking Eq.(\ref{gf}) would appear to suggest that this term
cannot modify the boundary conditions, in contradiction with what
we have just derived. 
However, with a general function ${\cal F}$ and in particular for 
$\beta {\cal R}$, the term
${d/ds} (l_a{\cal F}^{ab} \, t_b) \Phi$ 
appearing in its normal variation (see Eq.(\ref{eq:delii})) 
will be non-vanishing, and it is no longer appropriate to
discard a total derivative as we did in deriving Eq. (\ref{eq:delii}). 
For consistency, we claim therefore that we need to modify Eq.(\ref{gf})
as follows:
\begin{equation}
{\bf g} \to {\bf g}  - 
l_a {\cal F}^{ab}t_b \, {\bf n}\,.
\label{gp}
\end{equation}
For Gauss-Bonnet, the second term reads $-2\beta K_{\parallel\perp} {\bf n}$. 
This mysterious term is
precisely the tension associated with the edge energy 
$\oint_{\cal C} ds
\,\kappa$.
The projections of Eq.(\ref{gf}) along ${\bf l}$, ${\bf n}$ and ${\bf t}$ 
respectively then read
\begin{eqnarray}
&\alpha& [(K-K_0)^2  - 2 (K - K_0)  K_\perp] - 
2 \beta K_{\parallel\perp}^2 
\nonumber \\
&+& \sigma \kappa + \mu = 0 \,, \label{bg1}\\
&2& \alpha \nabla_\perp K  - 2\beta \dot{K}_{\parallel\perp} + \sigma
K_{\parallel} 
=0\,,\label{bg2}\\
&2& K_{\parallel\perp} [ \alpha (K - K_0) + \beta K_{\parallel} ] 
= 0 \,,\label{bg3}
\end{eqnarray}
where we have used the fact that, at the edge,
\begin{equation}
\dot {\bf n} = K_{\parallel} {\bf t} + K_{\parallel\perp} {\bf l}\,.
\end{equation}
As was the case in the previous section, these coincide with the
boundary conditions (\ref{bb1}), (\ref{bb2}), (\ref{bb3}) 
when $K_{\parallel\perp} \neq 0$.

\section{Conclusions}

Whereas
for a soap film, it is very simple to identify the forces operating on the edge, and so 
read off the boundary conditions on the bulk geometry, 
such an approach is less obvious for a  membrane. However,
we have demonstrated how simple geometrical and variational 
arguments may be exploited to 
derive the boundary conditions on the lipid membrane geometry.
We have made no restrictive assumptions 
about the symmetry of the configuration. 
We then showed how these boundary conditions emerge from
a balance of forces projected along a basis of vectors adapted to the edge.

We plan at a future date to examine the axially symmetric 
shape equation subject to the Gauss-Bonnet modified boundary conditions.
Matching observed 
configurations with solutions of the shape equation in this
more general setting provides a method, at least in principle, for determining the 
values of the two bulk rigidity moduli.

\begin{acknowledgments}
RC is partially supported by CONACyT under grant 32187-E.
JG and JS are partially supported by CONACyT 
under grant 32307-E and DGAPA at UNAM under grant IN119799.
\end{acknowledgments}

%\bibliographystyle{apsrev}
%\bibliography{strings}

\end{document}